\documentclass[useAMS]{mn2e}
\usepackage{lscape}
\usepackage{rotating}
\usepackage{amssymb}
\usepackage{float}

\def\msun{\hbox{M$_\odot$}}

\def\frac{\hbox{f$_{\rm mix}$}}

\def\t4{\hbox{t$_{\rm 4}$}}

\def\cm3{\hbox{cm$^{-3}$}}

\input psfig.sty
\input epsf.sty
\usepackage{graphicx}
\usepackage{epsfig}
\title[Extended Main Sequence Turnoff in NGC 1850]
{A Young Cluster With an Extended Main Sequence Turnoff: Confirmation of a Prediction of the Stellar Rotation Scenario}
\author[Bastian et al.] {N. Bastian$^{1}$, F. Niederhofer$^{2,3,4}$, V. Kozhurina-Platais$^{4}$, M. Salaris$^{1}$, S. Larsen$^{5}$, \newauthor I. Cabrera-Ziri$^{1,6}$,  M. Cordero$^{7}$, S. Ekstr\"om$^{8}$, D. Geisler$^{9}$, C. Georgy$^{8}$, M. Hilker$^{6}$, \newauthor N. Kacharov$^{10}$, C. Li$^{11}$,  D. Mackey$^{12}$, A. Mucciarelli$^{13}$, I. Platais$^{14}$ \\
$^{1}$Astrophysics Research Institute, Liverpool John Moores University, 146 Brownlow Hill, Liverpool L3 5RF, UK\\
$^{2}$Excellence Cluster Origin and Structure of the Universe, Boltzmannstr. 2, D-85748 Garching bei M\"unchen, Germany\\
$^{3}$Universit\"ats-Sternwarte M\"unchen, Scheinerstra\ss e 1, D-81679 M\"unchen, Germany\\
$^{4}$Space Telescope Science Institute, 3700 San Martin Drive, Baltimore, MD 21218, USA\\
$^{5}$Department of Astrophysics/IMAPP, Radboud University, P.O. Box 9010, 6500 GL Nijmegen, The Netherlands\\
$^{6}$European Southern Observatory, Karl-Schwarzschild-Stra\ss e 2, D-85748 Garching bei M\"unchen, Germany\\
$^{7}$Astronomisches Rechen-Institut, Zentrum f\"ur Astronomie der UniversitŠt Heidelberg, M\"onchhofstrasse 12-14, D-69120 Heidelberg, Germany.\\
$^{8}$Geneva Observatory, University of Geneva, Maillettes 51, 1290, Sauverny, Switzerland\\
$^{9}$Departamento de Astronomia, Universidad de Concepcion, Casilla 160-C, Chile\\
$^{10}$Max-Planck-Institut f\"ur Astronomie, K\"onigstuhl 17, D-69117 Heidelberg, Germany\\
$^{11}$Department of Physics and Astronomy, Macquarie University, Sydney, NSW 2109, Australia\\
$^{12}$Research School of Astronomy and Astrophysics, Australian National University, Canberra, ACT 2611, Australia\\
$^{13}$Department of Physics and Astronomy, University of Bologna, Viale Berti Pichat 6/2, I-40127 Bologna, Italy\\
$^{14}$Department of Physics and Astronomy, Johns Hopkins University, 3400 North Charles Street, Baltimore, MD 21218, USA\\
}
\date{Accepted. Received; in original form}
\pagerange{\pageref{firstpage}--\pageref{lastpage}}
\pubyear{2015}
\begin{document}
\maketitle
\label{firstpage}
\begin{abstract}
We present Hubble Space Telescope photometry of NGC~1850, a $\sim100$~Myr, $\sim10^5$~\msun\ cluster in the Large Magellanic Cloud.  The colour magnitude diagram clearly shows the presence of an extended main sequence turnoff (eMSTO). The use of non-rotating stellar isochrones leads to an age spread of $\sim40$~Myr. This is in good agreement with the age range expected when the effects of rotation in MSTO stars are wrongly interpreted 
in terms of age spread.  We also do not find evidence for multiple, isolated  episodes of star-formation bursts within the cluster, in contradiction to scenarios that invoke actual age spreads to explain the eMSTO phenomenon. NGC~1850 therefore continues the trend of eMSTO clusters where the inferred age spread is proportional to the age of the cluster.  While our results confirm a key prediction of the scenario where stellar rotation causes the eMSTO feature, direct measurements of the rotational rate of MSTO stars is required to definitively confirm or refute whether stellar rotation is the origin of the eMSTO phenomenon or if it is due to an as yet undiscovered effect.

\end{abstract}

\section{Introduction}
\label{sec:intro}

With high precision photometry now regularly accessible with the Advanced Camera for Surveys (ACS) and the Wide Field Camera 3 (WFC3) onboard the Hubble Space Telescope, stellar clusters have been studied in ever increasing detail.  This precision has allowed unexpected features in the colour magnitude diagram of young ($<1$~Gyr) and intermediate age ($1-2$~Gyr) clusters to be found and studied in detail (e.g., Mackey \& Broby Nielsen~2007).  One such feature is that the main sequence turn-off (MSTO) of massive Magellanic Cloud clusters is more extended than expected from a simple stellar population (i.e. a single isochrone) even when including the effects of photometric uncertainties and stellar binarity.  The origin of the extended MSTOs (eMSTOs) is still debated in the community.

While age spreads of the order of $200-700$~Myr appear to be the simplest explanation to the observed MSTO spreads (e.g, Mackey et al.~2008; Milone et al.~2009), such a scenario makes basic predictions that are at odds with observations.  The first is that massive clusters should show ongoing star-formation for the first few hundred Myr of their lives, whereas no clusters with ages beyond 10~Myr have been found with active star-formation (Bastian et al.~2013, Cabrera-Ziri et al.~2014; 2016a; Niederhofer et al.~2015a).  Moreover, In order to form a second generation of stars, clusters must be able to retain and/or accrete material from their surroundings (e.g., Conroy \& Spergel~2011).  However, clusters appear to be gas free after $2-3$~Myr, independent of their masses from $\sim10^4 - \sim10^7$~\msun\ (Bastian, Hollyhead, Cabrera-Ziri~2014; Hollyhead et al.~2015) and remain gas free for at least the next few hundred Myr (Bastian \& Strader~2014; Cabrera-Ziri et al.~2015; Longmore~2015).


An alternative explanation for the eMSTO phenomenon is that stellar rotation, which affects both the observational properties of the stars as well as their lifetimes through rotational mixing, can cause the spreads (Bastian \& de Mink~2009).  While initial works cast doubt on this mechanism (Girardi et al.~2011), stellar models that include rotation have been developed in recent years (e.g., Ekstr\"om et al.~2012, Georgy et al.~2013) which can be directly compared against observations.  Such comparisons have been done for intermediate age clusters (Brandt \& Huang~2015a,b) and it seems that for realistic rotational distributions (like those observed in open clusters - Huang et al.~2010) extended MSTOs are expected to be a common feature.  Niederhofer et al.~(2015b - hereafter N15b) extended this analysis to younger clusters ($<300$~Myr) and used the SYCLIST models (Georgy et al.~2014) to investigate the role of stellar rotation in affecting the MSTO\footnote{N15b effectively adopted a spread in rotation rates from $\omega=0-0.5$, where $\omega$ is the rotation rate divided by the critical rotation rate, independent of age.}.  They found that if a rotational distribution is present, but is interpreted incorrectly as an age spread, then the inferred age spread is proportional to the cluster age, with younger clusters showing smaller spreads.  Additionally, the models predict that, due to rotational mixing, there should be large ($\gtrsim0.5$~dex) star-to-star [N/H] variations that should correlate with position in the CMD.

N15b compared these predictions to observations of young and intermediate age clusters and found excellent agreement, with the inferred age spread expected to be $\sim30-40$\% of the age of the clusters.  The predictions nicely fit the observations of the young massive cluster NGC~1856 ($\sim300$~Myr), as interpreted by Milone et al.~(2015) and Correnti et al.~(2015).  One of the predictions of this interpretation is that if younger clusters were studied in the same way, that their inferred age spreads would be correspondingly shorter.

An excellent candidate to test this theory is NGC~1850, a $\sim100$~Myr, $2\times10^5$~\msun\ cluster in the Large Magellanic Cloud (Fischer et al.~1993; Niederhofer et al.~2015a).  If the rotational scenario is correct, then this cluster is expected to host an extended main sequence with an inferred age spread of $30-40$~Myr.  If the eMSTO phenomenon is caused by actual age spreads, due to for example its high escape velocity (e.g., Goudfrooij et al.~2014) then this cluster would be expected to show clear evidence of multiple bursts and/or a continuous star-formation history over its lifetime.  In the present work we use new Hubble Space Telescope observations to study the MSTO of NGC~1850 and compare it to predictions of the rotational scenario and the age spread scenario.  The paper is organised as follows: in \S~\ref{sec:observations} we present the observations and analysis tools used while in \S~\ref{sec:analysis} we search for spreads in the MSTO and compare it to expectations.  In \S~\ref{sec:discussion} we discuss our results and present our conclusions.

\section{Observations and Data Reduction}
\label{sec:observations}


NGC~1850 was observed as part of GO-14069 (PI-N. Bastian) with WFC3/UVIS camera onboard the Hubble Space Telescope through the F336W, F343N and F438W  filters with long, intermediate and short exposures. Photometry was carried out on the WFC3/UVIS images that were corrected for the imperfect CTE and simultaneously calibrated for bias, dark, low-frequency flats and the most recent UVIS zero-points (Rayn et al. in preparation).  Stellar photometry was derived with PSF fitting, using the spatially variable ``effective PSF" (ePSF) method (private communications, J. Anderson), with routines similar to that of ACS/WFC (Anderson \& van der Marel~2010).  The stellar positions were corrected for the WFC3/UVIS distortion (Bellini et al.~2011). Zeropoints were taken from the STScI website and aperture corrections were derived using isolated stars in the images.  More details regarding the photometry will be given in a forthcoming paper (Niederhofer et al. in prep.).


In order to subtract the background field stars, we followed the same procedure as used in Niederhofer et al.~(2015a, 2016).  We adopt a cluster field, centred on the cluster, with a radius of $2*r_{\rm c}$ (where $r_{\rm c} = 11.1"$ - McLaughlin \& van der Marel~2005), and a reference field  located near the edge of the images (i.e. as far from the cluster as possible) with the same area as the cluster field.  After defining the cluster and the reference field, we constructed CMDs for both fields. For every star in the reference field CMD, we removed the star in the cluster CMD that is closest to the reference field star in colour/magnitude space.  After extensive testing we found that this method better removes stellar contamination than applying a grid to the cluster/reference CMDs and subtracting stars from within grid cells.  This is discussed in detail in Cabrera-Ziri et al. (2016b).

\begin{figure}
\centering
\includegraphics[width=7cm, angle=-90]{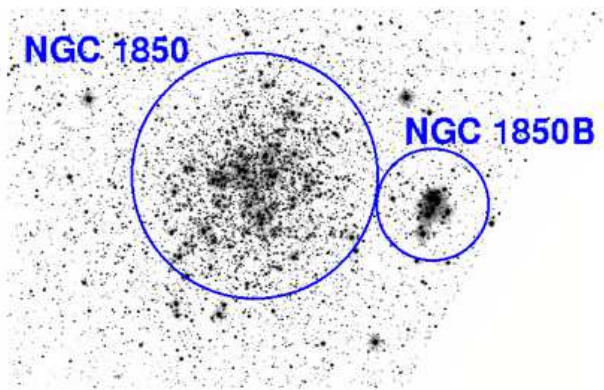}
\caption{$F438W$ image of NGC~1850 and NGC~1850B.  The circles denote the regions used in the present study (22" and 10" for NGC~1850 and 1850B, respectively). }
\label{fig:image}
\end{figure}

Additionally, there is a younger cluster, NGC~1850B ($\sim5$~Myr), nearby our primary target (see Fig.~\ref{fig:image}), NGC~1850 (Robertson~1974; Gilmozzi et al.~1994).  This cluster contributes stars to our 'cluster CMD' that were not subtracted based on our background method.  In order to remove these stars we defined in a first step an area with a radius of 10'' centred on NGC1850B and subtracted the background population using a field of the same area that is located opposite of NGC1850 with the same distance from the centre of the main cluster.  Then we used this background subtracted field of NGC 1850B to subtract the stars of this younger cluster from our science field.

For the analysis we adopt the BaSTI isochrones\footnote{http://www.oa-teramo.inaf.it/BASTI} (Pietrinferni et al. 2004, 2006) at $Z=0.008$ (McLaughlin \& van der Marel~2005).  The models have scaled solar abundances and include core overshooting.  These models do not include stellar rotation.

The extinction and distance modulus are found by comparing the isochrones to the shape/location of the main sequence fainter than F438W$=20$ (i.e. the curve in the lower MS which allows us to break any degeneracies between extinction and distance modulus.).  We adopt the same extinction coefficients as Milone et al.~(2015), namely $5.10*E(B-V)$ and $4.18*E(B-V)$ for the F336W and F438W filters, respectively.  We find best fit values of $E(B-V)=0.1$ and a distance modulus of $18.35$, consistent with previous work (e.g., Niederhofer et al.~2015a).

We have estimated the amount of differential extinction in the NGC~1850 field using the technique of Milone et al.~(2012).  As the extinction vector is largely parallel to the main sequence (MS; $18.5 < F438W < 20.5$) we used the lower part of the MS.  While the errors are larger in this region of the CMD, we found no significant differential extinction in the field.

\begin{figure*}
\centering
\includegraphics[width=5.5cm, angle=0]{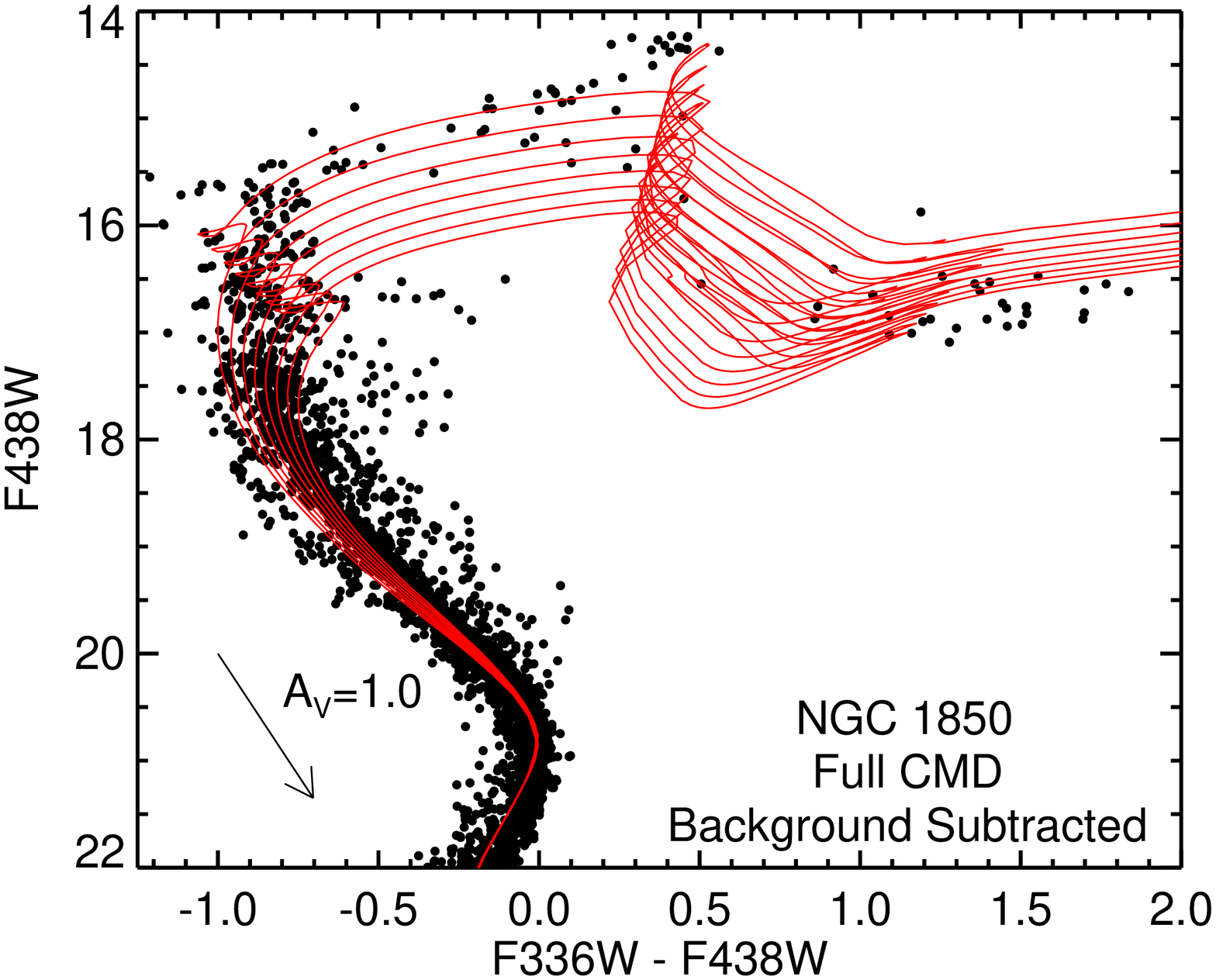}
\includegraphics[width=5.5cm, angle=0]{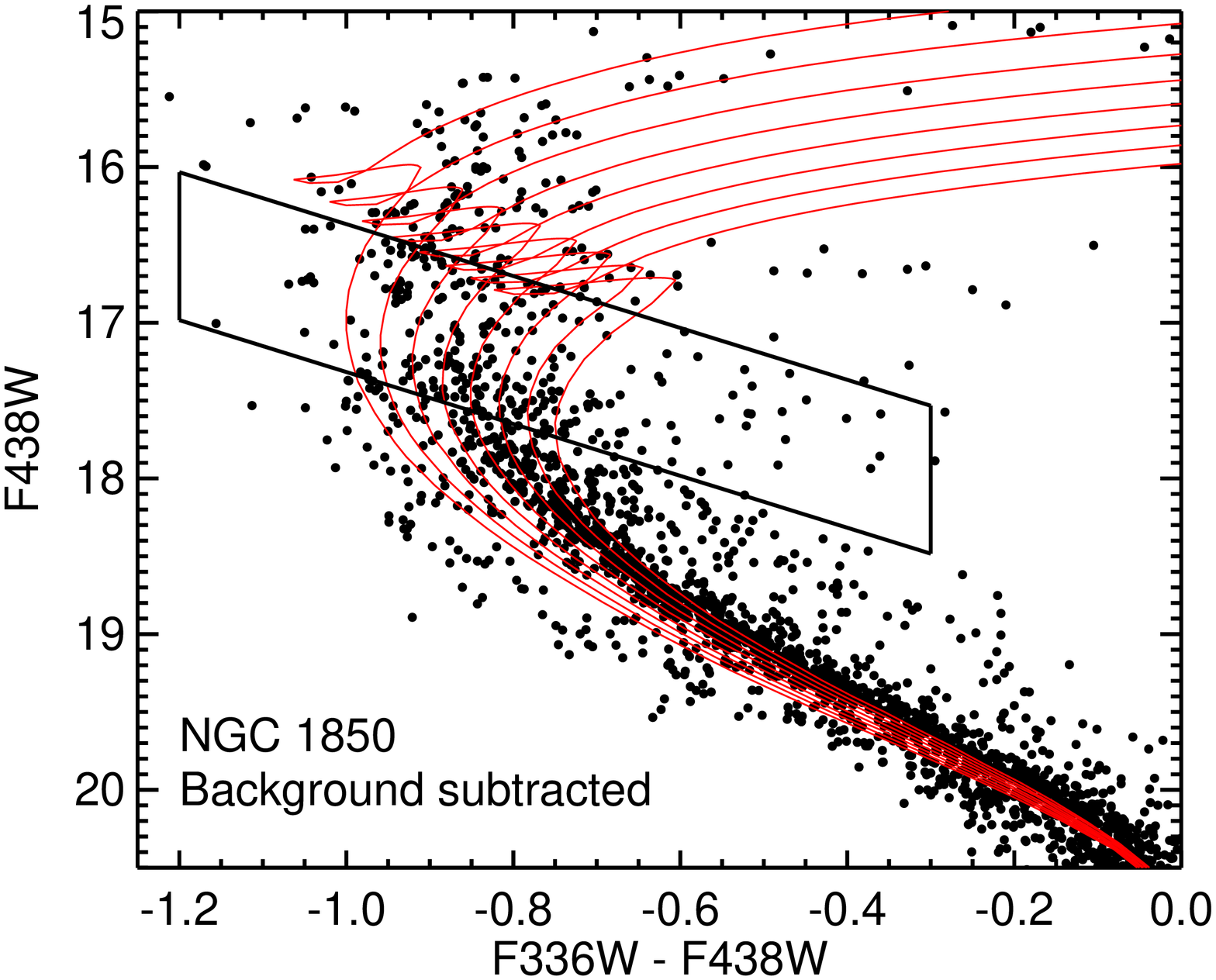}
\includegraphics[width=5.5cm, angle=0]{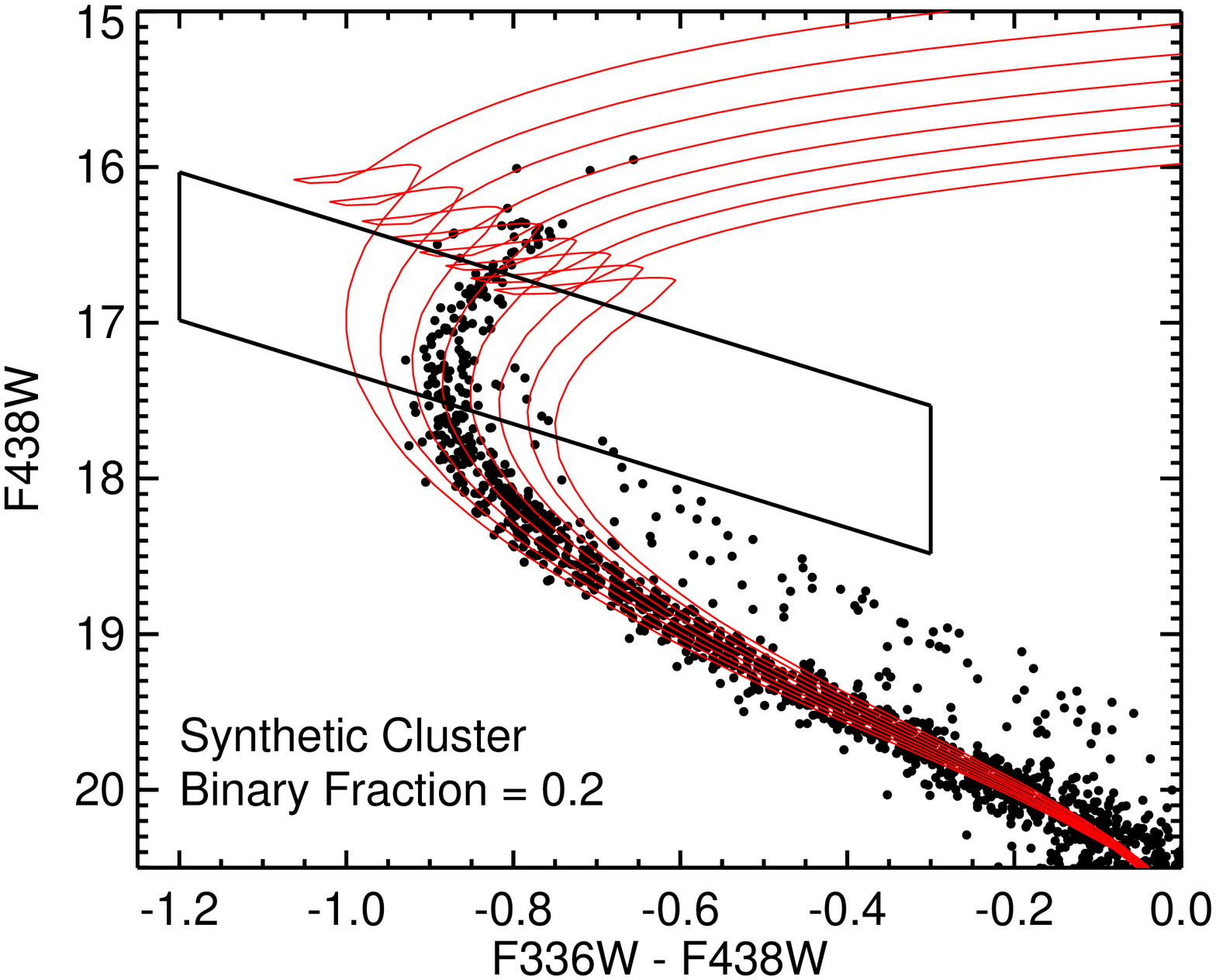}

\caption{{\bf Left panel:} Field star subtracted colour-magnitude diagram of NGC~1850.  We also show the BaSTI isochrones for Z=0.008 and a distance modulus of 18.35, for ages between 70 and 140~Myr in steps of 10~Myr.  {\bf Middle panel:} A zoom in on the main sequence turn-off (MSTO) portion of the CMD showing that despite the tight lower sequence, the MSTO is clearly extended. {\bf Right panel:}  A synthetic cluster with no age spread, errors taken from the observations, a total binary fraction of 20\% (flat mass ratio distribution).  This simulation was used to determine the expected age spread from binarity and photometric errors. The cut across the MSTO used in the analysis is shown in the middle and right panels.}
\label{fig:cmd}
\end{figure*}

We show the observed, background subtracted, extinction corrected CMD of NGC~1850 in Fig.~\ref{fig:cmd}.

In order to estimate the binary fraction within the cluster we created synthetic clusters, based on the BaSTI isochrones and a Salpeter~(1955) stellar IMF, with photometric uncertainties taken from the observations.  We adopt an age of $100$~Myr and a flat mass ratio distribution (in agreement with other works on young massive clusters - e.g., Milone et al.~2015), and only included binaries with $q>0.5$ (where $q$ is the mass ratio of the secondary to the primary), and varied the binary fraction, $f_{\rm bin}^{q>0.5}$.  We then verticalised the main sequence between $18.5 < F438W < 20.0$ in the observations and synthetic clusters and made a histogram of sources in colour.  The histograms were normalised to the same number of stars, and the synthetic cluster most closely resembling the observations was selected.  We found the best fit was obtained for $f_{\rm bin}^{q>0.5} = 0.1-0.15$, which if we extrapolate to the full mass ratio distribution leads to $f_{\rm bin}^{total} = 0.2-0.3$.  This synthetic cluster is shown in the right panel of Fig.~\ref{fig:cmd} and we will use it as a comparison cluster in the subsequent analysis.

\section{Analysis}
\label{sec:analysis}

\begin{figure}
\centering
\includegraphics[width=8cm, angle=0]{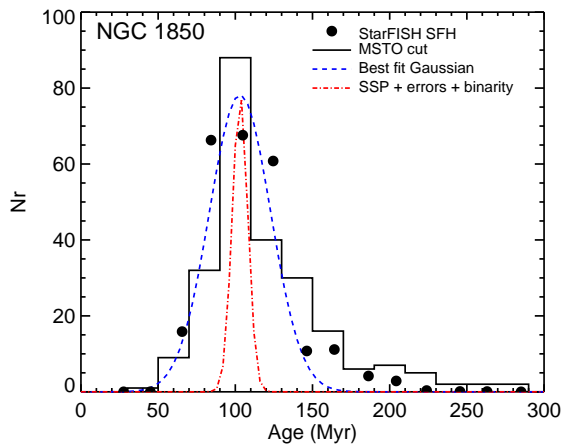}

\caption{The inferred age spread across the MSTO of NGC~1850.  The histogram shows the observed distribution (see text for details) whereas the blue dashed line shows the best Gaussian fit to the data.  The red dashed line (narrower distribution) shows the expected spread due to photometric errors and stellar binarity.}
\label{fig:sfh}
\end{figure}

As can clearly be seen in the centre panel of Fig.~\ref{fig:cmd}, the CMD of NGC~1850 displays a main sequence turn-off that is more extended than would be expected from photometric errors and/or binarity if the underlying population was a simple stellar population.  In order to quantify the spread we have carried out two experiments.  First, we took a cut across the MSTO, perpendicular to the isochrones and assigned an age to each star in this box based on its proximity to the nearest isochrone.  This is similar to what was done in Goudfrooij et al.~(2014; hereafter G14), Li et al.~(2014; 2016) and Bastian \& Niederhofer~(2015).  The resulting age distribution is shown in Fig.~\ref{fig:sfh} as a histogram.  The dashed blue line represents the best fitting Gaussian to the histogram.  In order to estimate the intrinsic spread due to photometric errors and binarity, we carried out the same fitting procedure on the synthetic cluster discussed in \S~\ref{sec:observations} and shown in the right panel of Fig.~\ref{fig:cmd}.  The distribution was then fit with a Gaussian function, which is shown as the red dashed line in Fig.~\ref{fig:sfh}.

The MSTO of NGC 1850 is significantly more extended than expected for a single isochrone, including the effects of photometric errors and binarity.  The best fitting Gaussian has a dispersion of $\sim20$~Myr.  The expected dispersion for an SSP (including errors and binarity) is $\sim6$~Myr.  If we subtract the SSP result from the observational result of NGC~1850 in quadrature, we find that the intrinsic dispersion is $18.9$~Myr, corresponding to a FWHM of $44$~Myr \footnote{We note that the inferred age spread induced by the photometric uncertainties and binaries is less than the age difference between consecutive isochrones ($10$~Myr), so the dispersion of $6$~Myr is only an estimate.  If we would use instead a $10$~Myr spread, the corrected eMSTO spread of NGC~1850 would decrease to $40$~Myr.}.

We have also fit the observed CMD with the star-formation history (SFH) fitting StarFISH package (Harris \& Zaritsky 2001), similar to what was done in Niederhofer et al.~(2016).  We only fit the MSTO portion of the CMD, namely ($14 \le F336W \le 19$ and $-1.4 \le F336W-F438W \le 0.0$), and we adopted a flat binary distribution with $f_{\rm bin}^{total} = 0.25$.  The results are shown in Fig.~\ref{fig:sfh} as filled circles and we have normalised the StarFISH distribution to have the same area under the curve as the MSTO distribution.  Overall the agreement with the SFH derived from the cut across the MSTO is quite good.

As noted in previous works on the intermediate age clusters (Li et al.~2014; 2016; Niederhofer et al.~2016), the post-main sequence distribution of stars (e.g., the SGB)  in some clusters does not appear to be consistent with an age spread within the cluster, instead being narrower and also concentrated towards the 'young' end of the distribution expected from the MSTO  (although see Goudfrooij et al.~2015 for an alternative view).  The SGB does appear to be narrow in the SYCLIST models including rotation at this age, but this will be investigated in more detail in a future work.

\section{Discussion and Conclusions}
\label{sec:discussion}

\begin{figure}
\centering
\includegraphics[width=8cm, angle=0]{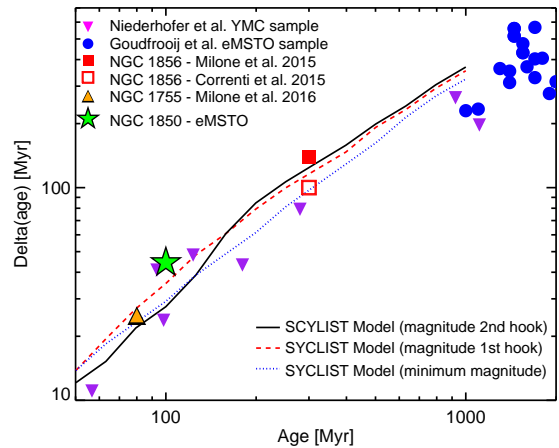}

\caption{The inferred age spread in a sample of YMCs and intermediate age clusters.  NGC~1850, based on the extended MSTO, is shown as a star.  We also show predictions from the Geneva models with rotation (Georgy et al.~2013) of the inferred age spread expected if a rotational spread is mis-interpreted as an age spread.  Note the strong correlation between the age of the cluster and the inferred age spread, in good agreement with the model predictions.  Adapted from N15b.}
\label{fig:delage}
\end{figure}

The CMD of NGC~1850 shows evidence for an extended MSTO, similar in nature to that found for the older NGC~1856 ($\sim300$~Myr; Milone et al.~2015; Correnti et al.~2015) and the $1-2$~Gyr intermediate age clusters in the LMC/SMC (e.g., Mackey \& Broby Nielsen~2007; G14), however the inferred age spread is significantly smaller than in the other clusters studied to date.  In Fig.~\ref{fig:delage} we show our results for NGC~1850 in the cluster age vs. inferred age spread plane (N15b), shown as a filled star.  The points shown, except the YMC Niederhofer et al.~(2015a) sample, were all determined in a similar way, i.e. finding the FWHM of the inferred age distribution across the MSTO.  NGC~1850 continues the trend reported by Niederhofer et al.~(2015b; 2016) that the age spread inferred for young and intermediate age clusters is directly proportional to the age of the cluster itself.  N15b predicted that, based on its age of $\sim100$~Myr, NGC~1850 should 1) display an extended MSTO and 2) that the inferred age spread (when analysed with non-rotating stellar isochrones) should be $\sim30$~Myr, in good agreement with the measured $44$~Myr.


The observations presented here are not consistent with the previously suggested interpretation that the eMSTO phenomenon is due to true age spreads within the cluster, as the correlation between cluster age and the inferred age spread is not expected in such a scenario.  G14 have suggested a limit in the escape velocity, above which ($\sim10-15$~km/s) clusters can retain the ejecta of AGB stars and form a second generation of stars (this scenario also requires large amounts of gas to be accreted from the surroundings).  We note that the lack of abundance spreads within the eMSTO clusters is inconsistent with this scenario (e.g., Mucciarelli et al.~2014).  The current escape velocities of many clusters that host MSTO spreads are well below the proposed $10-15$~km/s limit (e.g., G14; Milone et al.~2016; Piatti \& Bastian~2016), hence G14 assume that clusters  begin their lives with much higher masses and escape velocities, and would lose stars due to tidal effects and also the cluster would expand during it's lifetime.  The assumptions behind this calculation, such as the applicability of the adopted model and the resulting extreme mass loss, have been discussed and questioned elsewhere (Niederhofer et al.~2016; Cabrera-Ziri et al.~2016a).

However we can use NGC~1850 to test this scenario directly.  At the age of NGC~1850, in the G14 scenario, the cluster would not have been expected to have lost much of its initial mass yet, so the derived age distribution should be representative of the initial distribution.  Hence, if we applied the G14 scenario to NGC~1850 we would not expect to observe a single Gaussian distribution in age, but rather a large peak when the first generation formed, followed by a smaller (Gaussian) peak due to the formation of the 2nd generation.  This is clearly at odds with the observations, which show a single Gaussian type distribution.  The same conclusion can be drawn from the observed age distribution of the $\sim300$~Myr cluster, NGC~1856 (Milone et al.~2015; Correnti et al.~2015).

The inferred age distribution of NGC~1850, being well approximated by a single (possibly skewed) Gaussian distribution, is the same form seen in the majority of clusters studied to date which host extended MSTOs, although there are notable exceptions, e.g., NGC~1846 (Mackey et al.~2008) and NGC 1783 (Rubele et al.~2013 - although see Niederhofer et al.~2016) which show bi-modal distributions.  If stellar rotation is the underlying cause of the eMSTOs, it would imply a specific rotational distribution (potentially bi-modal) for the stars in these clusters (see also D'Antona et al.~2015 and Milone et al.~2016).

Comparison of the observed CMDs of young and intermediate age clusters with stellar models that include rotation (e.g., Brandt \& Huang~2015b; N15b; D'Antona et al.~2015) suggest a relatively good agreement between the two, offering support to the notion that stellar rotation is the cause of the eMSTO phenomenon (Bastian \& de Mink~2009).  However, a definitive test of this scenario will be to measure the rotational velocity ($V_{\rm rot}sin(i)$) of a large sample of stars across the eMSTO.  This is potentially feasible in the intermediate age clusters, although due to the faintness of the MSTO at this age in the LMC/SMC this will push current instrumentation to its limits.  On the other hand, younger clusters such as NGC~1850 and NGC~1856, offer an excellent chance to obtain rotational velocities for large samples of stars along the MSTO due to their increased brightness at this age.  Another crucial test of the rotational scenario will be to measure chemical abundances of stars across the MSTO, which are expected to be affected by rotational mixing, especially [N/H] (e.g., Georgy et al.~2013).

In summary, NGC~1850 is another young massive cluster that hosts an extended main sequence turn-off.  If the eMSTO is interpreted as an age spread, based on non-rotating stellar models, the inferred age spread is $\sim40$~Myr.  This is in good agreement with predictions of the rotational scenario, i.e., if a spread due to a distribution of rotation rates (like that observed in open clusters - Huang et al.~2010) was misinterpreted as an age spread (Niederhofer et al.~2016).  The inferred age distribution is also not in agreement with the scenario of actual age spreads put forward by G14, as in such a case the inferred age spread would be expected to be made up of two separate star-formation events, instead of the single smooth Gaussian type distribution found here.

\vspace{-0.7cm}
\section*{Acknowledgments}

NB gratefully acknowledges financial support from the Royal Society (University Research Fellowship) and 
the European Research Council (ERC-CoG-646928, Multi-Pop).  D.G. gratefully acknowledges support from the Chilean 
BASAL Centro de Excelencia en Astrof\'isicay Tecnolog\'ias Afines (CATA) grant PFB-06/2007.  Support for this project was provided by NASA through grant HST-GO-14069 from the Space Telescope Science Institute, which is operated by the Association of Universities for Research in Astronomy, Inc., under NASA contract NAS5Ð26555.

\bsp
\label{lastpage}
\end{document}